\documentstyle[preprint,aps]{revtex}
\begin{document}
%\draft
\title{ \bf
LATTICE $(\Phi^4)_4$ EFFECTIVE POTENTIAL GIVING \\
\vspace*{3mm}
             SPONTANEOUS SYMMETRY BREAKING AND\\
THE ROLE OF THE HIGGS MASS }

\author{ A. Agodi, G. Andronico and M. Consoli}

\address{ Dipartimento di Fisica - Universita di Catania; \\
 Istituto Nazionale di Fisica Nucleare - Sezione di Catania;\\
Corso Italia, 57 - I 95129 Catania - Italy}
\newcommand{\beq}{\begin{equation}}
\newcommand{\eeq}{\end{equation}}

\maketitle

\begin{abstract}
We present a critical reappraisal of the available results on the broken
phase of $\lambda(\Phi^4)_4$ theory, as obtained from rigorous formal
analyses and from lattice calculations.
All the existing evidence is compatible with Spontaneous Symmetry Breaking
but dictates a trivially free shifted field that becomes
controlled by a quadratic hamiltonian in the continuum limit.
As recently pointed out, this implies that
the simple one-loop effective potential should become effectively
exact. Moreover, the usual naive assumption that the Higgs mass-squared
$m^2_h$ is proportional to
its ``renormalized'' self-coupling
$\lambda_R$ is not valid outside perturbation theory: the appropriate
continuum limit has $m_h$ finite and vanishing $\lambda_R$.
A Monte Carlo lattice computation of the $\lambda(\Phi^4)_4$ effective
potential, both in the single-component and in the O(2)-symmetric cases,
is shown to agree very well with the one-loop prediction. Moreover,
its perturbative leading-log improvement (based on the concept of
$\lambda_R$) fails to reproduce the Monte Carlo data.
These results, while supporting in a new fashion the peculiar ``triviality''
of the $\lambda(\Phi^4)_4$ theory, also
imply that, outside perturbation theory, the magnitude of
the Higgs mass does not give a measure of the observable interactions in
the scalar sector of the standard model.

\end{abstract}
\vskip 15 pt
\widetext

\par  {\bf 1.Introduction}

The so called effective potential of a quantum field
theory is a powerful tool to get informations about the structure
of its vacuum, whose analysis, in turn, gives insight into its
physical content. A reliable determination of the effective potential
requires a critical control of details not directly accessible
to a perturbative calculation. A convenient approach to uncovering the
non perturbative features of such details may start by trying to
determine the continuum limit of some discrete version of the theory
on a lattice. In this approach one exploits the correspondence of an
Euclidean (imaginary-time) quantum field theory to a relativistic
(real-time) theory, the Schwinger's functions of the former determining
the Green's functions of the latter.
While substantial experience has been already gained in the numerical
analysis of lattice field theories, the reliability of statements
about their continuum limits is still open to questions, due to the
fact that the equivalence of different procedures to get those limits
is not yet controlled by theoretically well defined conditions. This
amounts to a
shift of the problem from the domain of bare computation, including
the evaluation of numerical errors and/or approximations, to that of
interpretation, exploiting the connections of the numerical results
with the formal theory or with suitable models of the continuum limit.
\par

The continuum $\lambda(\Phi^4)_D$ theories of a self-interacting scalar
field $\Phi$ in $D$ spacetime dimensions are perhaps the simplest
examples to study as limits of their versions on a lattice.
Their physical interest stems from their role in the
Standard Model, as Spontaneous Symmetry Breaking (SSB) and the Higgs
phenomenon \cite{higgs} determine
the masses of the vector bosons and some other essential features
of the Electroweak Sector of the model \cite{WS}, e.g. with $\Phi$
being a (non hermitian) isospin doublet.\par

The scalar $\lambda(\Phi^4)_D$ theories are characterized by a rather
peculiar $D$-dependence of their dynamical content: they have been proved
to be "trivial" ( i.e. equivalent to a generalized free field theory) for
$D > 4$, but not so ( i.e. truly interacting) for $D < 4$
\cite{froh,glimm,book,latt}.
For $D = 4$ the situation is a bit less definite, since, instead of a
proof,
one has several ``evidences''(based, of course, on different methods
of analysis, whose assumed correctness points out a lack of equivalence)
suggesting, in some sense, either ``triviality'' or ``nontriviality'', the
latter being, admittedly, an almost ignored option \cite{ped}.\par

 The commonly accepted evidence about the ``triviality'' of
$\lambda(\Phi^4)_4$ is frequently assumed to support
the view that the scalar sector of the standard model,
dynamically generating the vector boson masses through the Higgs mechanism
\cite{higgs}, can only be an effective theory, valid up to some cutoff scale.
Without a cutoff, the
argument goes, there would be no scalar self-interactions, and without them
there would be no symmetry breaking \cite{call,lang}.  This point
of view also leads to upper bounds on the Higgs mass \cite{call,lang,neu}.

  Recently \cite{resolution,zeit}, it has been pointed out, on the
basis of very general arguments, that SSB is not
incompatible with ``triviality'', at least in the case of the scalar
field theory with quartic self-interactions.
One could have a non-zero vacuum expectation value (VEV) of the field,
$\langle\Phi\rangle$, yet
find only non-interacting, quasi-particle excitations above the vacuum.
This picture, while providing a consistent model-theory characterized
by the coexistence of ``triviality'' and of a non-trivial effective potential
\cite{cian,return,cast,bran,con,new,iban,u1,rit}, also clarifies the meaning
of some  {\it explicit} studies of ``triviality'' in the { \it broken}
phase ($\langle\Phi\rangle \ne 0$) \cite{broken}.

As discussed in \cite{resolution,zeit},
since the effective potential determines the zero-momentum 1PI vertices of the
symmetric phase \cite{CW}, its structure might carry some traces of
non trivial interactions modifying the
zero-momentum-mode sector of the symmetric phase in a fashion which
somehow de-stabilizes the symmetric vacuum and gives rise to SSB.

This remark \cite{resolution,zeit},
immediately brings out the relevance of studying the {\it massless}
$\lambda(\Phi^4)_4$ theories, as, for them, the zero-momentum ($p^{\mu}=0$) is
a physical on-shell point (inaccessible to perturbation theory).

If, in the continuum limit, {\it all} interaction
effects of the symmetric phase can be reduced into a set of Green's
functions expressible in terms of just the
first two moments of a Gaussian (functional) distribution, then one gets
a description
of SSB consistent with ``triviality''
which does not contradict any analytical or numerical result
\cite{cian,return,cast,bran,con,new,iban,u1,rit,ped}.
In this case the shifted field $h(x)=\Phi(x)-\langle\Phi\rangle$
becomes governed by a quadratic Hamiltonian and
 the simple ``one-loop'' potential turns out to be effectively {\it exact}.
 More precisely,
 the one-loop potential and the gaussian effective potential
determine the same relation
\beq
         m^2_h=8\pi^2v^2_R
\eeq
 between the physical
Higgs mass $m_h$ and the physical field VEV $v_R$.
This turns out to be true as well in the post-gaussian
approximations (see, in particular, \cite{rit2}), where the
Higgs propagator $G(x,y)$ is properly optimized
at each value of $\phi=\langle\Phi\rangle$, by solving the corresponding
non-perturbative gap-equation $G=G_o(\phi)$
\cite{resolution,zeit,con,new,iban,rit,rit2}.
This result follows from
the exact connection of the effective potential $V_{\rm eff}(\phi)$ with
the effective potential for composite operators $E[\phi,G]$ \cite{CJT}
$$   \int d^3x~ V_{\rm eff}(\phi)=E[\phi,G_o(\phi)]      $$
and from ``triviality'', which entails that all interaction effects
(apart from those included in the physical vacuum structure) should be
reabsorbed into the physical mass of the ``generalized free field'' $h(x)$
defined above.
In any approximation consistent with the (assumed) exact ``triviality''
of the theory the mass of the Higgs field cannot represent a
measure of the observable interactions in the broken phase which, in the
absence of interactions involving other fields, should contain only
trivially free excitations.
\par  As we shall explain in sect. 2, this simple picture
of an interacting zero-momentum mode that produces a non-trivial shape
for the effective potential, giving SSB, co-existing
with trivially
non-interacting excitations in the broken phase,
is not in contradiction with any knowledge derived from both
{\it all} formally rigorous theorems about $\lambda(\Phi^4)_4$ in the
broken phase and {\it all} available lattice calculations.

It is well kown, since a long time, that the effective potential of
$\lambda(\Phi^4)_D$ can be analysed in a rather simple way on a lattice
by computing the VEV of the field as a function of a (global or uniform)
"external source" strength $J$ \cite{call2,huang,huang2} .\par

We report, in sects.3,4, an analysis of the data obtained from
precise Monte Carlo lattice calculations of the effective potential
for both the single component
and the O(2)-symmetric $\lambda(\Phi^4)_4$ theory.

We have performed such an analysis for several values of the bare
quadratic ($r_o$) and quartic ($\lambda_o$) couplings in the Euclidean
discrete action in four dimensions
\beq
   S = \sum_x [{{1}\over{2}}\sum_{\mu}(\Phi_{x+\mu} - \Phi_x)^2 +
{{r_o}\over{2}}\Phi^2_x  +
{{\lambda_o}\over{4}} \Phi_x^4 - J \Phi_x ]
\eeq
where $x$ stands for a general lattice site and, unless otherwise stated,
lattice units are understood.\par
The numerical results from a version of
Metropolis' Monte Carlo algorithm have been best fitted with analytical
forms possibly consistent with "triviality" and allowing for some
suitably chosen ``corrections''.\par

As a matter of fact,
the one-loop potential agrees remarkably well with the lattice results,
while its
perturbative ``improvement'' fails to reproduce the Monte Carlo data.\par
Thus, definite support is obtained for the picture proposed in
\cite{resolution,zeit}.
The basic physical interest of our analysis, emphasizing a
rather unconventional but theoretically not surprising peculiarity of
``renormalization'' in this Bose field example, also stems from the new
interpretation given to the Higgs mass, no more appearing as proportional
to a renormalized coupling constant (whose vanishing is required by
``triviality''),
with remarkable new implications on the dynamics of the Higgs particle
in the Standard Model.\par

As we shall explain in sect.2, our numerical results agree with those
already obtained from lattice calculations, as far as we know, and our
theoretical analysis is fully consistent with the general conditions imposed
on the theory by the coexistence of SSB and triviality.\par

Our analysis also shows that ``triviality'' and perturbation theory, within any
attempt to construct a cutoff-independent and {\it non-vanishing} renormalized
coupling at non zero external momenta $\lambda_R$,
are in inner contradiction \cite{trivpert}.
 The (new) evidence of this incompatibility is given
by the lattice computation of the effective potential which turns out
to represent just
the sum of the classical potential and of the zero-point energy
of a {\it free} field. Outside perturbation theory, $m_h$ and $\lambda_R$
are not proportional and while the latter vanishes the former remains an
arbitrary parameter. This result has important physical
implications as discussed in sect.5 where we shall also propose our
conclusions.
\vskip 25 pt
\par  {\bf 2. SSB and ``triviality''}
\par In our study of SSB the basic quantity to
compute is the VEV of the bare scalar field $\Phi(x)$
\beq
      \langle \Phi \rangle _J = \phi_B(J)
\eeq
in the presence of an ``external source'' whose strength $J(x)=J$ in Eq.(2)
is $x$-independent.\par
In fact, determining $\phi_B(J)$
at several $J$-values is equivalent \cite{huang,huang2} to inverting the
relation
\beq
        J=J(\phi_B)={{dV_{eff}}\over{d\phi_B}}
\eeq
involving the effective potential $V_{eff}(\phi_B)$. Actually, starting from
the action in Eq.(2), the effective potential of the theory
can be {\it rigorously} defined, up to an arbitrary integration constant,
through the above procedure.

In this framework, the occurrence of SSB is determined
by exploring (for $J\neq0$) the properties of the function
\beq
               \phi_B(J)=-\phi_B(-J)
\eeq
in connection with its behaviour in the limit of vanishing $J$
\beq
  \lim_{J\to 0^{\pm}}~\phi_B(J)=\pm v_B \neq 0
\eeq
over a suitable range of the bare parameters
$r_o,\lambda_o$ appearing in the lattice action Eq.(2).

The existence of the $\lambda(\Phi^4)_4$ {\it critical point} can be
stated, in the framework connecting the lattice theory with its continuum
limit, on the basis of rigorous formal arguments \cite{glimm}. Namely,
for any $\lambda_o > 0$, a critical value $r_c=r_c(\lambda_o)$ exists,
such
that, for $r_o < r_c$, one finds non trivial solutions of Eq.(6). {\it After}
fixing the vacuum and hence $\langle\Phi\rangle=v_B$, one can define
a shifted field
\beq
            h(x)=\Phi(x)-v_B
\eeq
whose VEV vanishes by construction.
\par As reviewed in the last section of \cite{book}, the following
statements hold true. For $D > 4$, the continuum limit, precisely
defined in Ch.15 of \cite{book},
is ``trivial'' in the sense that the connected n-point Green's functions
($n\geq 3$)
vanish in that limit. Moreover, the shifted field renormalization
constant $Z_h$ (as derived from the renormalization of the term containing the
field {\it derivatives} in the bare action) must be {\it finite}. As a
consequence, if one {\it assumes} a single renormalization constant
for both the field VEV $v_B$ and the shifted field $h(x)$,
one cannot get both
a finite Higgs mass and a finite vacuum expectation value in the
continuum limit \cite{book}.
Indeed, the bare field VEV $v_B$ becomes  {\it infinite} in units of
the mass $m_h$ of the shifted field which, in this specific context,
must be trivially free.

In this picture, where the same ``$Z$''
controls the scaling properties of {\it both} vacuum field and fluctuations,
namely
\beq
            Z=Z_h
\eeq
what amounts to agree that
\beq
         v^2_R={{v^2_B}\over{Z_h}}
\eeq
the above quoted statements \cite{book} requiring
\beq
 {{m^2_h}\over{v^2_B}} \to 0
\eeq
and the finiteness of $Z_h=Z$, also entail
\beq
 {{m^2_h}\over{v^2_R}}
        ={{m^2_hZ_h}\over{v^2_B}}\sim
 {{m^2_h}\over{v^2_B}} \to 0
\eeq
In this approach, where ${{v^2_B}\over{Z_h}}\sim v^2_B$ is kept fixed and
related to the Fermi constant, $m^2_h$ vanishes in the continuum limit and
one concludes that the theory makes sense only in the presence of an
ultraviolet cutoff. This point of view, also leads to upper bounds on the
Higgs mass \cite{call,lang,neu}.
In all interpretations of the lattice simulations performed
so far, the validity of Eq.(8) has been {\it assumed}. Let us refer
to the very complete review by C.B.Lang \cite{lang}. There, the value of
$Z=Z_h$ is extracted from the large distance
decay of the Goldstone boson propagator and used in Eq.(9) to define
a renormalized field VEV $v_R$ from the bare field VEV $v_B$ as
measured on the lattice.
 Since the lattice data both support the trend in Eq.(10) and provide
 $Z_h=1$ to very good accuracy (see Tab. II of \cite{lang})
one deduces (11).
The limiting behaviour (11), obtained in this way,
is usually interpreted (see \cite{lang} Fig.19, caption included)
to be a test of the validity of the perturbative relation
\beq
    m^2_h=\lambda_R v^2_R
\eeq
in which
the {\it renormalized} coupling $\lambda_R$, at external momenta
comparable to the Higgs mass itself, has been introduced.
\par As discussed in the Introduction, $\lambda_R$ is
a spurious concept in a ``trivial'' theory.
On the other hand, without assuming the validity of any perturbative
or quasi-perturbative relation, the lattice data confirm
the conclusions of \cite{book}, namely, in the continuum
limit
\par~~~~a) the bare field becomes infinite in units of the Higgs mass, i.e.
(10)
\par~~~~b) the shifted field (fields) is (are) trivially free since $Z_h\to 1$
\par Point a), supported by theoretical arguments and by the
results of lattice simulations, can be taken as a basic condition defining the
continuum limit of SSB in $\Phi^4_4$ theories.
\par Point b) is completely consistent with the usual treatment, based on
the Lehmann spectral representation of the propagator,
for a field with {\it vanishing} VEV. To this end
one identifies \cite{zimmermann} the $Z_h$ of the
continuum theory, related to the (integral of the) spectral function $\rho(s)$,
with the limit-value of
its cutoff-dependent form when the ultraviolet regulator is (continuously)
removed. Thus $Z_h=1$ implies $\rho(s)$ to be just a $\delta$-function,
since the relevant fields have a vanishing VEV.
\par Points a) and b) describe two basic outcomes of ``triviality'' in
the broken phase. A consistent description of SSB in the $\lambda(\Phi^4)_4$
theory
should explain the non-uniform scaling of the Higgs mass and of the bare
vacuum field while preserving, at the same time,
the non interacting nature of the shifted field(s).
\par Does
perturbation theory provide any clue to understand a) and b) ?
Strictly speaking,
the perturbative one-loop $\beta$-function, exhibiting an unphysical
Landau pole, does not define {\it any} continuum limit. The usually
accepted view maintains that the interval between the origin and
the Landau pole becomes arbitrarily large in the limit
$\lambda_R \to 0$, thus allowing to recover the continuum limit of the
theory in agreement with ``triviality'' while still preserving, at finite
ultraviolet cutoff, the validity of the two perturbative relations (9,12).
This interpretation of ``triviality'' can be questioned for the following
reasons. In perturbation theory, $\lambda_R$ is the basic Renormalization
Group (RG) invariant whose cutoff-independence generates the cutoff-dependence
of the bare coupling $\lambda_o=\lambda_o(a)$, with $a$ denoting the lattice
spacing equivalent to the ultraviolet cutoff $\Lambda$. Different $\lambda_R$'s
define different {\it theories}.
 It would be contradictory, and inconsistent with the original premise that
$\lambda_R$ is finite and cutoff-independent, to now conclude that
$\lambda_R $ has any implicit or explicit cutoff dependence that causes it
to vanish in the limit $a \to 0$.
 ``Triviality'' means that $\lambda_R$, which represents
a measure of the observable interactions, must vanish
identically in order to get a flow of the bare parameters
$r_o=r_o(a)$, $\lambda_o=\lambda_o(a)$ extending down to $a=0$.
 The one-loop
 perturbative $\beta$-function, obtained under the assumption $\lambda_R>0$
(and ${{9\lambda_o}\over{16\pi^2}}\ln {{\Lambda^2}\over{m^2_h}}<1$ )
 simply does not allow for such an extension. Furthermore, the above
perturbative interpretation of ``triviality'' is inconsistent with explicit
 two-loop calculations, where the perturbative $\beta$-function exhibits a
non trivial ultraviolet fixed point at non zero bare coupling
$\lambda_o=\lambda^*>0$ (let us remark that rigorous arguments \cite{froh}
can be given to qualify as ``spurious'' this additional fixed point).
In the perturbative approach, if $\lambda_R$ is taken to represent
a value of the bare coupling at some finite momentum scale,
$\lambda_o=\lambda_o(a)$ flows towards $\lambda^*$ when $a\to 0$ for
{\it any} ~$0<\lambda_R<\lambda^*$.
\par Moreover, the perturbative explanation
of Eq.(10), namely $m^2_h \to 0$ at $v^2_B\sim {{v_B}\over{Z_h}}=fixed$, is
certainly not
unique. One might as well consider, e.g. the alternative possibility
$v^2_B \to +\infty$, $ m^2_h=fixed$, if the renormalized vacuum field
$v_R$ and $v_B$ are non trivially
related through a renormalization constant $Z_{\phi}\neq Z_h$.
 This alternative,
first proposed in \cite{return} and later on thoroughly discussed
in \cite{resolution,zeit,rit2}, does not violate any basic principle.
 Let us emphasize that
 the vacuum field renormalization $Z_{\phi}$ has nothing to do, in principle,
with the cutoff-dependence of the term containing the field {\it derivatives}
in the bare action.
The usual assumption of a single ``field''-renormalization factor
derives from giving an operatorial meaning to the field rescaling, namely
$$        "~ \Phi(x)=\sqrt{Z}~\Phi_R(x)~ "              $$
This relation is a consistent shorthand for expressing the
``wave function''-renormalization in a theory allowing an asymptotic
Fock representation, but no more so
in the presence of SSB, since, then, it overlooks that the shifted field
is not defined before fixing the vacuum. The Lehmann
spectral-decomposition argument constrains only $Z_h$.
\par The previous discussion should make clear that the
perturbative interpretation (9,12) of the basic results a) and b) might
be hardly taken as correct.
At the same time, it should make apparent that the problems associated
with the peculiar structure of the continuum limit of the $\lambda(\Phi^4)_4$
theory must be faced anew with open minded critical attention.
\par In the following we shall briefly review
the nonperturbative
renormalization pattern which arises from the
analysis of the effective potential in those approximations which are
consistent with ``triviality'', as referred to in the Introduction and
explained in \cite{resolution,zeit}. This will clearly show how
the two basic results a) and b) are consistently derived in our approach.
The basic difference
with respect to perturbation theory, where one
attempts to define a continuum limit
in the presence of {\it observable} interactions,
lies in
the choice of the quantities that are required to remain finite in the
continuum limit:
namely, the vacuum energy and the ``gap'' in the energy
spectrum corresponding to the mass of the {\it free}-field quanta.
\par In terms of the bare vacuum field $\phi_B= \langle \Phi \rangle$ and
 of the bare coupling $\lambda_o$ we obtain the well-known result
for the one-loop effective potential \cite{CW}
\beq
 V^{{\rm 1-loop}}(\phi_B)  =  {{\lambda_o}\over{4}} \phi^4_B +
\frac{\omega^4(\phi_B)}{64\pi^2}
\left( \ln \frac{\omega^2(\phi_B)}{\Lambda^2} -
\frac{1}{2} \right)
\eeq
where, by definition, $\omega^2(\phi_B)=3\lambda_o\phi^2_B$
and $\Lambda$ denotes the Euclidean ultraviolet cutoff equivalent to the
lattice spacing ($\Lambda\sim {{\pi}\over{a}}$).
The minimum condition requiring
\beq
 m^2_h =\omega^2(v_B)  =
3 \lambda_o v^2_B =
\Lambda^2~\exp (-{{16\pi^2}\over{9\lambda_o}}).
\eeq
with $Z_{\phi}={{8\pi^2}\over{3\lambda_o}}$, also allows giving
the one-loop effective potential in the form
\beq
V^{\rm 1-loop}(\phi_B)
={{\pi^2\phi^4_B}\over{Z^2_{\phi} }}
(\ln{{\phi^2_B}\over{v^2_B}}-{{1}\over{2}}) ,
\eeq
Finally, the ground state energy is
\beq
W=V^{\rm 1-loop}(v_B)=-{{m^4_h}\over{128\pi^2}}
\eeq
By defining the physical vacuum field $\phi^2_R={{\phi^2_B}\over{Z_{\phi}}}$
in such a way that all zero-momentum interaction effects are
reabsorbed into its normalization, i.e. so that
\beq
   {{d^2V^{\rm 1-loop}}\over{d\phi^2_R}}|_{\phi_R=\pm v_R}=m^2_h
\eeq
holds true, we get \cite{resolution,zeit,con,iban,new}
\beq
V^{\rm 1-loop}(\phi_R)=\pi^2\phi^4_R
(\ln{{\phi^2_R}\over{v^2_R}}-{{1}\over{2}})
\eeq
and
\beq
 m^2_h=8\pi^2v^2_R
\eeq
The non-perturbative nature of the vacuum field
renormalization ($Z_{\phi}\sim 1/\lambda_o$), first discovered in the
gaussian approximation by Stevenson and Tarrach \cite{return}, should not be
confused with the $h$-field wave function renormalization. In fact, at one
loop, $h$ is just a free field with mass $\omega(\phi_B)$, and hence
 one has trivially $Z_h=1$, in agreement with the basic result b). The
structure with
$Z_{\phi}\neq Z_h$, is allowed by the Lorentz-invariant nature of the field
decomposition into $p_{\mu}=0$ and $p_{\mu}\neq 0$ components
\cite{resolution,zeit}. It is, of course, more general than the one
assumed in perturbation theory, and its unconventional appearance
is, in a sense, an indication of the peculiar non perturbative nature
of the physical vacuum in the massless $\lambda(\Phi^4)_4$ theory,
which allows SSB to coexist with ``triviality''. Notice that the possibility
of a different rescaling for the vacuum field and the fluctuations is
somewhat implicit in the conclusions of the authors of
\cite{book} ( see their
footnote at page 401: "This is reminiscent of the standard procedure in the
central limit theorem for independent random variables with a nonzero mean: we
must subtract a mean of order $n$ before applying the rescaling $n^{-1/2}$ to
the fluctuation fields").
\par To further illustrate
 the asymmetric treatment of the zero-momentum mode ( which is well
known to play a special role in a Bose system at zero temperature) one can
consider the analogy with the quantum mechanical 3-dimensional
$\delta$- potential discussed by Huang \cite{huang2} and Jackiw \cite{beg}.
 In this example,
the exact solution of the Schr\"odinger equation
requires the radial wave-function to vanish at the origin. This condition is
automatically satisfied for all partial waves except S-waves. For S-waves it
cannot be satisfied if the wave function is required to be continuous at the
origin. In that case, there would be no S-waves, and the solutions
of the equation would not form a complete set. But a discontinuity at the
origin might be acceptable, since the potential is singular there. By
defining the $\delta$-potential as the limit of a sequence of square well
potentials, it is straightforward to show that, in the limit of zero-range, the
S-wave solution vanishes discontinuously at the origin. Not surprisingly,
all scattering phase shifts vanish. The analogy with $\lambda\Phi^4$ theory is
remarkable. There is no observable scattering, and yet the particle is not
completely free. It is free everywhere except at the origin where the S-wave
vanishes abruptly. In our field theoretical case, all finite momentum modes
are free and as such there is no non-trivial S-matrix. However, the
unobservable zero-momentum mode is a truly dynamical entity whose non-trivial
self-interaction is not in contradiction with "triviality" being
 reflected in the zero-point energy of the free shifted field.
\par In conclusion, the basic result a) is
in complete agreement with the analytical results derived from
 the one-loop effective potential. Indeed,
the finiteness of the RG-invariant ground state energy in Eq.(16) requires
that the Higgs mass has also to be finite.
In the continuum limit one gets a cutoff-independent $m_h$
and a finite ground state energy in connection with the
vanishing of the bare coupling as
\beq
    \lambda_o={{16\pi^2}\over{9\ln {{\Lambda^2}\over{m^2_h}} } }
\eeq
In this limit, from Eq.(14), one rediscovers, with $Z_h=1$
identically, the trend in (10)
 in complete agreement with the previous points a) and b).
Eqs.(18.19) also
 hold in the more sophisticated gaussian and
postgaussian approximations \cite{resolution,zeit,con,new,rit2}. It is
important to note that the logarithmic trend in
Eq.(20), controlling the cutoff dependence of the ratio
${{m^2_h}\over{v^2_B}}$ (see Eq.(14) ), allows to explain the observed
logarithmic decrease on the lattice of the ratio
$${{m^2_hZ_h}\over{v^2_B}}\sim {{m^2_h}\over{v^2_B}}\sim
{{1}\over{ \ln {{\Lambda^2} \over{m^2_h}} }}    $$
(see fig.19 of \cite{lang}) without any need
of introducing the leading-log $\lambda_R$
$$    \lambda_R={{\lambda_o}\over
{1+{{9\lambda_o}\over{16\pi^2}}\ln{{\Lambda^2}\over{m^2_h}}}}  $$
Indeed, the usual perturbative interpretation of ``triviality'' (see fig.6 of
\cite{lang} ), based on the assumed reliability of the above equation
at $\lambda_o=+\infty$, i.e.
$$    \lambda_R={{16\pi^2}\over{9\ln {{\Lambda^2}\over{m^2_h}} } },    $$
neglects that the leading-log resummation is not defined in that region
and, as previously recalled, contradicts explicit perturbative two-loop
calculations.
\vskip 10 pt
\par  {\bf 3. Lattice computation of $\phi_B(J)$.}
\par The VEV of the bare
field $\langle \Phi \rangle_J = \phi_B(J)$ is the simplest quantity to
compute on the lattice, and its $J$-dependence can be exploited to get
the first derivative of the effective potential, Eq.(4), as already
recalled at the beginning of sect.2. In this sense,
a computation of the slope of the effective potential, based on the
response of the system to an applied external source, has the advantage of
being fully {\it model-independent}.
 At the same time, since it is well known \cite{glimm,call2} that the
data become unstable for small values of the external source, we have to
restrict our analysis to a "safe" region of $J$ (in our case $|J|\geq 0.05$)
 where finite size effects appear to be negligible as checked on several
lattices (see below).
Indeed, at  smaller $|J|$
 the precision of the data becomes uncertain
(not only but especially in the neighborhood of the critical point)
and this was confirmed by testing the accuracy
of the data against Eq.(5). The loss of accuracy observed in the
neighborhood of $J = 0$ cannot be accounted for by numerical and
statistical errors but is sensitive
to finite size effects.
In this way we have been led to avoid
 any ``direct'' lattice calculation of $\phi_B(J)$
at $J = 0$, at least in a ``small sized'' lattice as $10^4$, and
this is the main motivation for our choice of a method not
relying on such ``direct'' calculations in the analysis of the
lattice data.
\par In order to compare the lattice data with the continuum theory one has
to employ some definite model. As recalled in the Introduction, the
equivalence of different procedures to get this limit is not yet safely
controlled by theoretically well defined conditions.
 For instance, as discussed
by Lang \cite{lang} one can use an effective theory (the so called
"chiral perturbation theory")
to relate the lattice quantities computed in a finite volume
at non-zero $J$ to those at $J=0$ in an infinite volume. The cutoff-dependence
of such ``thermodynamic'' quantities can be then compared with models of
the continuum limit.
\par In our case, we shall
use two basically different functional forms, based on
 the one-loop effective potential and its
perturbative leading-log improvement, to fit the measured values of $\phi_B(J)$
and to extrapolate $\phi_B(J)$ toward $J\to 0^{\pm}$.
Our choice is motivated by the discussion of SSB and ``triviality''
presented in sect.2.
However, anybody could use our data,
 collected in a "safe" region of $J$ and determined in a
model-independent way, to compare with his own preferred model
and draw the appropriate conclusions.
\par The calculation of $\phi_B(J)$,
on the basis of the action given in Eq.(2), has been performed by
using a version of the Metropolis' Monte Carlo algorithm with periodic boundary
conditions, for several values of the parameters $r_o, \lambda_o, J$
appearing in the discrete action. Other currently used algorithms, as the
cluster method and the hybrid Langevin's approach,
are presently under our investigation and the results
will be compared in
a forthcoming paper. However, possible improvements
are only expected in the region of very small $J$
which is not analyzed here.
\par The calculations have been carried out on DEC 3000 AXP (Mod.400
and 500) computers.\par
With the aim of obtaining a reliable control on the precision of
the data, we have carefully analyzed their dependence on the choice of the
random number generator, of the ``thermalization'' process and
of the dimension of the sample (number of ``sweeps'' or ``iterations'')
entering the
determination of the relevant data. The results for $J=J(\phi_B)$
obtained from runs
on our $10^4$ lattice have been compared with those reported in
\cite{huang,huang2} and the agreement with the plots $J=J(\phi_B)$ at
$\lambda_0 = 1$ in Fig.2 of \cite{huang} (suitably enlarged) has been
found better than $1\%$ in the range $|J|\geq 0.05$. Other comparisons with
\cite{huang,huang2} have not been attempted in view of their direct
determination of $v_B$ at $J=0$ and of their assumption $Z_{\phi}=Z_h$ in the
evaluation of the shifted field propagator.
\par
After 60,000 iterations,
 which correspond to the data reported in Table I, the calculated
$\phi_B(J)$ is stable at the level of the first three significant digits
for all $|J|$-values. This stability has been controlled with several
runs up to 200,000 iterations. As $|J|$ increases, the stability gets better
(down to the level of $10^{-4}$ in the explored range of $J$-values)
and finite size effects appear negligible, as
checked by comparing with the runs on a $16^4$ lattice
and, subsequently, confirmed by an independent computation performed
by P. Cea and L. Cosmai \cite{cea} reported below.
\par In order to introduce our analysis of the Monte Carlo data, let us first
discuss the predictions from the one-loop potential which, as discussed in
sect.2, reflect the ``triviality'' of the theory.
 By differentiating Eq.(15), we obtain the
``bare source''
\beq
J(\phi_B)={{dV^{\rm 1-loop}(\phi_B)}\over{d\phi_B}}=
{{4\pi^2\phi^3_B}\over{Z^2_{\phi} }}\ln{{\phi^2_B}\over{v^2_B}} ,
\eeq
which can be directly compared
 with the lattice results for $J=J(\phi_B)$.
Strictly speaking, just as the effective potential
is the convex envelope of Eq.(15) \cite{convex,rit},
Eq.(21) is valid only for $|\phi_B|>v_B$ and
$J(\phi_B)$ should vanish in the
presence of SSB in the range $-v_B \leq \phi_B \leq v_B$. This means that
 the average bare field $ \phi_B(J)=\langle\Phi_B\rangle_J$,
while satisfying the relation (5), i.e.
$$             \phi_B(J)=-\phi_B(-J)~    \eqno(5)  $$
for any $J\neq 0$,
should tend to the limits stated in Eq.(6). However,
as discussed above,
Eq.(5) is not well reproduced numerically
for $|J|\sim 0.01$ or smaller and
we shall have to restrict our analysis to a ``safe'' region
of $J$-values ( $|J| \geq 0.05$) allowing to get
the values of $v_B$ and $Z_{\phi}$ from a best fit to the lattice
data by using Eq.(21) after having identified on the
lattice the ``massless'' regime, i.e. the one corresponding to a
renormalized theory
with no intrinsic scale in its symmetric phase $\langle \Phi \rangle=0$.
\par Usually, this would require to find the value of the quadratic bare
coupling $r_o$ for which \cite{CW}
$${{d^2V}\over{d\phi^2_B}}|_{\phi_B=0}=0 $$
 However, the region around $\phi_B=0$
being not directly accessible, we have argued as follows.
 We start with the general expression \cite{zeit}
\beq
J(\phi_B) = \alpha \phi_B^3 \ln(\phi_B^2/v_B^2) +
\beta v^2_B \phi_B (1 - \phi_B^2/v_B^2),
\eeq
 which is still consistent with ``triviality'' (corresponding to an effective
potential given by the sum of a classical background and the zero point
energy of a free field) but allows for an explicit
scale-breaking $\beta$-term. Setting
$\alpha=0$ one obtains a good description of the data in the ``extreme double
well'' limit ($r_o$ much more negative than $r_c$,
where $r_c$ corresponds to the onset of SSB)
where SSB is a semi-classical phenomenon and the
zero-point energy represents
a small perturbation. Then we start to increase $r_o$, at
fixed $\lambda_o$, toward the unknown value $r_c$ and look at
the quality of the fit with $(\alpha,\beta,v_B)$ as free parameters. The
minimum allowed value of $r_o$
 such that the quality
of the 2-parameter fit $(\alpha,\beta=0,v_B)$ is exactly the same as in the
more general 3-parameter case will {\it define} the ``massless'' case so that
 we can fit the data for $J=J(\phi_B)$ with our Eq.(21).
Finally, Eq.(14) suggests that the vacuum field $v_B$, in lattice units,
 vanishes extremely fast when $\lambda_o \to 0$.
 Thus, to avoid that noise and signal become comparable,
 we cannot run the lattice simulation
at very small values of $\lambda_o$ but have to restrict to values
$\lambda_o \sim 1$
such that still
${{\lambda_o}\over{\pi^2}}<<1$ but $v_B$, in lattice units, is not too
small. Smaller values of $\lambda_o$ ($\sim$0.4-0.6),
however, should become accessible with the
largest lattices available today ($\sim100^4$) where one should safely
reach values $|J|\sim$0.001 being still in agreement with Eq.(5).
 At $\lambda_o=1$
we have identified the massless regime at a
 value  $r_o=r_s<r_c$ where $r_s \sim -0.45$.
By using the accurate weak-coupling
relation between the bare mass and the euclidean cutoff \cite{CW}
\beq
 r_s=-{{3\lambda_o}\over{16\pi^2}}\Lambda^2
\eeq
and using the relation $\Lambda=\pi y_{Q}$
(where $y_{Q}$ is expected to be O(1))
we obtain $y_{Q}\sim 1.55$.
 It should be noted that our operative definition of the ``massless'' regime
amounts to determine for weak bare coupling the
numerical coefficient relating ultraviolet cutoff and lattice spacing for
quadratic divergences ($\Lambda \sim 4.87$ ).
This agrees well with an independent analysis
of the L\"uscher and Weisz
lattice data presented by Brahm \cite{bra}.
In fact Brahm's result
(in the range $\lambda_o \leq 10$ )
is $\Lambda =4.893\pm0.003$.
 Also, ref.\cite{bra} predicts the massless regime to correspond
to $r_s=-(0.224\pm0.001)$ for
$\lambda_o=0.5$ in the infinite-volume limit. This implies
$r_s\sim-(0.448\pm0.002)$ for $\lambda_o=1$,
in excellent agreement with our result $r_s\sim-0.45$. Thus, the
model-dependence in the definition of the massless regime, introduced by our
fitting procedure to Eq.(22), is negligible for
${{\lambda_o}\over{16\pi^2}}<<1$.
\par Finally, the massless relation (14) predicts,
\beq
  (v_B)^{\rm 1-loop}=
{{\pi y_{L}}\over{\sqrt{3\lambda_o }}} \exp(-{{8\pi^2}\over{9\lambda_o}})
\eeq
and
\beq
Z^{\rm 1-loop}_{\phi}={{8\pi^2}\over{3\lambda_o}}.
\eeq
In Eq.(24) we have used $y_{L}$ rather than $y_{Q}$ since one
does not expect precisely the same coefficient to govern
the relation between euclidean cutoff and lattice spacing for both
quadratic and  logarithmic divergences.
\par The consistency of Eqs.(24,25) with the corresponding values obtained
from the best fit to the lattice data checks the validity of the picture
given in sect.2 for the coexistence of SSB and ``triviality''. We have
determined $y_L$ at
$\lambda_o=1$ and, after that, the scaling behaviour of $v_B$ with the bare
coupling has been directly compared with Eq.(24). On the other hand, the
prediction in Eq.(25) for $Z_{\phi}$, which fixes the absolute normalization
of the curve $J=J(\phi_B)$, only depends on $\lambda_o$ and can be
directly compared with the lattice data in the massless regime.

\par Our numerical values for $\phi_B(J)$ and the
results of the 2-parameter fits to the data by using Eq.(21)
 for $\lambda_o=$ 0.8,1.0 and 1.2 are shown in Table I together with the
one-loop prediction (25). The change  of $r_o$
with $\lambda_o$ is computed
 by using Eq.(23) and our result $r_sa^2\sim -0.45$ for $\lambda_o=1$.
The reported errors are due to three main sources :
\par~~~a) the purely statistical errors from the Monte Carlo sampling of the
field configurations
\par~~~b) errors which can be easily recognized as numerical artifacts (in our
case they can be extimated by comparing with Eq.(5) )
\par~~~c) finite size effects which we have extimated by computing a few points
on a $16^4$ lattice
\par By fixing $Z_{\phi}$ to its one-loop value (25) and using Eq.(24) for
$v_B$ we have determined the value
$y_L=2.07\pm 0.01$
( ${{\chi^2}\over{d.o.f.}}={{6}\over{21}}$)
which
 corresponds to $v_B=(5.814\pm0.028)10^{-4}$ at $\lambda_o=1$. Once we
know $y_L$ we can predict the value of $v_B=(v_B)^{\rm Th}$ and compare
with the results of the fit to the lattice data at
$\lambda_o$=0.8 and 1.2 . The results in Table II
show a remarkable agreement of the Monte Carlo data with
the one-loop predictions.
\par As an
additional check that our values for $\phi_B(J)$ are
not affected by appreciable systematic effects, we have compared with a
completely
independent Monte Carlo simulation performed by P. Cea and L. Cosmai \cite{cea}
 on a
$16^4$ lattice. Their data for $\lambda_o=1$ and $r_o=-0.45$ are reported
in Table III. Note that their errors are statistical only and are smaller by
an order of magnitude than those reported in our Table I (where errors of type
b) and an estimate of the errors of type c) were included). Still, the quality
of the one-loop fit is excellent confirming that finite size effects do not
affects the results of the fits.
\par At this point
one may ask: "How does the standard
perturbative approach compare with the lattice data ?".
This is not a trivial question. Indeed, the same lattice data might be
consistent with different functional forms thus allowing for different
extrapolations at $J=0$ and, hence, different interpretations.
In particular, since our data are collected in the region of small bare
coupling (${{\lambda_o}\over{\pi^2}}<<1$ ), it is reasonable to check
whether the lattice data
 agree better with the one-loop potential
or with its perturbative leading-log improvement, related
to the concept of $\lambda_R$. This is even more important in our case
because:
\par~~~i) The leading-log approximation provides a commonly accepted
 perturbative interpretation of ``triviality'' (ignoring the objections
raised in sect.2)
\par~~~ii) Just the validity of
 the leading-log resummation is usually considered as an indication
that the one-loop minimum cannot be trusted and there is no SSB in the
massless regime \cite{CW}. If this were true the correct estrapolation at
$J=0$ would require $v_B=0$
\par However, one should be aware that
the higher order corrections to the one-loop
potential represent genuine self-interaction effects of the shifted field which
we know from ``triviality'' {\it cannot} be
physically present in the continuum limit.
Were we observing the need for perturbative
corrections of the one-loop formula this would
contradict all the non perturbative evidence about $h(x)$ being a generalized
free field.
\par By fitting our lattice data to the
leading-logarithmically ``improved'' version of Eq.(21), namely
\[
 J^{{\rm LL}}(\phi_B)={ {\lambda_o\phi^3_B}\over{1+{{9\lambda_o}
\over{8\pi^2}}
\ln{ { \pi x_{{\rm LL}} }\over{|\phi_B|}} }}
\]
($x_{{\rm LL}}$ denoting an adjustable parameter for each
$\lambda_o$) we find, for 21 degrees
of freedom,
 $(\chi^2)_{{\rm LL}}=53,163,365$ for $\lambda_o=0.8,1.0,1.2$
 respectively (to compare with the values $\chi^2=$5,6,13 of the corresponding
one-loop 1-parameter fits). The same comparison
with the Cea-Cosmai data reported in Table III gives the result
$$         x_{LL}=1.627\pm0.001            $$
$$   {{\chi^2}\over{\rm d.o.f}}={{11100}\over{9-1}}   $$
 This should be compared with the one-loop results
 ${{\chi^2}\over{d.o.f.}}={{3.2}\over{9-2}}$ reported in Table III for the
2-parameter fit and the value
 ${{\chi^2}\over{d.o.f.}}={{3.3}\over{9-1}}$ for the 1-parameter fit when
$Z_{\phi}$ is constrained to its one-loop value $Z_{\phi}=26.319$.
These results show that the agreement between lattice data and one-loop
formulas are
not a trivial test of perturbation theory but, rather, a non
perturbative test of ``triviality''. They also verify the consistency of
our definition of the continuum limit presented in sect.2.

\par In order
to fully appreciate what the above results exactly mean we have reported
in Tables IV and V
the two theoretical sets $J=J^{Th}(\phi_B)$ (Th=one-loop, leading-log)
which best fit
 the measured values of $\phi_B$ (with their errors $\Delta J$ corresponding to
the error in $\phi_B$). Finally, in the last column, we report the quantity
$\chi^2={{(J^{\rm{EXP}}-J^{\rm{Th}})^2}/{(\Delta J)^2}}$ where
$J=J^{\rm{EXP}}$ is the "experimental", input value of $J$ at which the various
$\phi_B$ have been computed by Cea and Cosmai.
By inspection of Tables IV and V one finds:
\par~~~ 1) the
leading-log formula is accurate to the level of a few percent. The fit routine
tries to reproduce the lattice data with a suitable value of the free
parameter $x_{LL}$ within the logarithm, thus optimizing the
relation between euclidean cutoff and lattice spacing and effectively
reabsorbing also non-leading perturbative corrections.
 In this case the
experimental data and the leading-log curve intersect each other somewhere in
between $J$=0.5 and $J=0.4$.
However, their slopes are clearly different as it can be seen
from the fact that the theoretical prediction is lower than the experimental
data for $J>0.5$ and
higher for $J<0.4$. Since the Monte Carlo data are so precise the value
of the $\chi^2$ is enormous.
\par ~~~2) the one-loop prediction is accurate to a completely different level,
namely O($10^{-4}$). This result cannot be understood from the conventional
perturbation theory viewpoint; apparently a cruder approximation
reproduces the
lattice data remarkably well, while its "pro forma" improvement fails badly !
However, if one understands our interpretation
of ``triviality'' , then this is
precisely what one expects.
\par In conclusion,
the lattice computations nicely confirm the conjecture
\cite{resolution,zeit} that, as a consequence of ``triviality'', the one-loop
potential becomes effectively exact in the continuum limit.
Indeed, there is a well defined region in the space of the
bare parameters $(r_o,\lambda_o)$,
controlled by Eq.(23), where the effective
potential is described by its one-loop approximation to very high accuracy.
The lattice analysis of the effective potential in
 this region, i.e. close to the critical line and in the broken
phase, where the theory is {\it known} to become gaussian, should be
pursued with larger lattices and, possibly, also with different methods.
 Indeed,
by increasing the lattice size, our analysis can be further extended towards
the physically relevant point ${{\lambda_o}\over{\pi^2}}\to 0^+$,
$r_o\to 0^-$ which
corresponds, in the continuum limit of the theory, to
``dimensional transmutation'' \cite{CW} from the classically scale invariant
case. Our results confirm the point of view \cite{resolution,zeit} that
the massless version
 of the $\lambda(\Phi^4)_4$ theory, although ``trivial'', is not ``entirely
trivial'': it provides, at the
same time, SSB and
 a meaningful continuum limit $\Lambda \to \infty$, $\lambda_o \to 0^+$ such
that the mass of the free shifted field in Eq.(14) is cutoff independent.
However, the mass $m_h$ does not represent a measure of any observable
interaction since the corresponding field $h(x)$ is described by a quadratic
hamiltonian.

\vskip 25 pt
\par  {\bf 4.The O(2)-symmetric case}
\par As
discussed in \cite{con,new,resolution}, and explicitly shown in \cite{rit2},
 one expects Eqs.(18,19), of the
one-component theory,
 to be also valid for the radial field in the O(N)-symmetric case.
 This observation
originates from ref.\cite{dj} which obtained, for the
radial field, the same effective potential as in the
one-component theory. This is extremely
intuitive. The Goldstone-boson fields contribute to the
effective potential only through their zero-point energy, that is
an additional constant, since, according to ``triviality'', they are
free massless fields.
Thus, in the O(2)-symmetric case, one may take
 the diagram $(V_{eff},\phi_B)$ for the one-component theory
and ``rotate'' it around the $V_{eff}$ symmetry axis. This generates a three-
dimensional diagram $(V_{eff},\phi_1,\phi_2)$ where $V_{eff}$, owing to
the O(2) symmetry, only depends on the
bare radial field,
\beq
                \rho_B=\sqrt{\phi^2_1+\phi^2_2}
\eeq
in exactly the same way as $V_{eff}$ depends on $\phi_B$ in the one-component
theory; namely
($\omega^2(\rho_B)=3\lambda_o\rho^2_B$)
\beq
 V^{{\rm 1-loop}}(\rho_B)  =  \frac{\lambda_o}{4} \rho^4_B +
\frac{\omega^4(\rho_B)}{64\pi^2}
\left( \ln \frac{\omega^2 (\rho_B) }{\Lambda^2} -
\frac{1}{2} \right) .
\eeq
By using Eqs.(14,25) $V^{\rm 1-loop}$
 can be re-expressed in the form
\beq
V^{\rm 1-loop}(\rho_B)={{\pi^2\rho^4_B}\over{Z^2_{\phi} }}
(\ln{{\rho^2_B}\over{v^2_B}}-{{1}\over{2}}) ,
\eeq
\par By differentiating Eq.(28), we obtain the bare ``radial source''
\beq
J(\rho_B)={{dV^{\rm 1-loop}(\rho_B)}\over{d\rho_B}}=
{{4\pi^2\rho^3_B}\over{Z^2_{\phi} }}\ln{{\rho^2_B}\over{v^2_B}} ,
\eeq
which we shall compare
 with the lattice results for $J=J(\rho_B)$.
\par The lattice simulation of the O(2)-invariant theory has been obtained,
with periodic boundary conditions, from the action
\beq
\sum _x~[ {{1}\over{2}}{ \sum^4_{\mu=1}\sum^2_{i=1}(\Phi_i(x+e_{\mu})-
\Phi_i(x))^2}
+{{1}\over{2}} r_o\sum^2_{i=1}(\Phi_i^2(x))
 + \frac{\lambda_o}{4}( \sum^2_{i=1}\Phi^2_i(x))^2
- \sum^2_{i=1}J_i\Phi_i(x)]
\eeq
where $\Phi_1$ and $\Phi_2$ are coupled to two constant external sources
$J_1$ and $J_2$.
By using $J_1=J\cos\theta$ and $J_2=J\sin\theta$ it is straightforward to show
that, having defined $\phi_{1}=\langle\Phi_{1}\rangle_{J_1,J_2}$ and
$\phi_{2}=\langle\Phi_{2}\rangle_{J_1,J_2}$, the bare radial field
\beq
          \rho_B=\sqrt{\phi^2_1+\phi^2_2}
\eeq
does only depend on $J$, that is
\beq
\rho_B=\rho_B(J)
\eeq
We started our Monte Carlo simulation on a $10^4$ lattice by investigating
first
the $(r_o,\lambda_o)$ correlation corresponding
to the classically scale-invariant case.
\par In the O(2) case, we have used
  the results of the previous section and of
\cite{bra} to get the analogous of Eq.(23)
for the two-component theory. Since we are considering
 ${{\lambda_o}\over{\pi^2}}<<1$,
this simply introduces a combinatorial factor of
$4/3$ so that Eq.(23) becomes
\beq
 r_s=-{{\lambda_o}\over{4\pi^2}}\Lambda^2
\eeq
Hence, we expect the massless case to correspond to $r_s\sim -0.6$ for
$\lambda_o=1$. This was confirmed
 to good accuracy by using the fitting procedure to the analogous of Eq.(22)
defined with $\rho_B$ in the place of $\phi_B$. Thus, the identification of the
massless regime on the lattice appears to obey the simple scaling laws
(23,33) and is under theoretical control.
\par Just like in the single-component case
 Eq.(5) was poorly
reproduced numerically at small values of $J$,
in the O(2)-symmetric case we find that, at small $J$,
the exact $\theta$-independence
of $\rho_B$ (see Eq.(32) )
is poorly reproduced and a ``direct'' data processing becomes unreliable.
We therefore consider a ``safe'' region
of $J$-values, $J \geq 0.05$, in which the spurious $\theta$-dependence
gives the data an uncertainty less than
$\pm 3\%$. Fitting the data to Eq.(29) we can infer the values
 of $v_B$ and $Z_{\phi}$ and compare with Eqs.(24,25).
\par Our numerical values for $\rho_B(J)$ in the massless case,
 obtained with our version of the Metropolis' Monte Carlo
algorithm on the $10^4$ lattice, are reported in Table VI for
$\lambda_o=$1.0, 1.5 and 2.0.
For each $\lambda_o$ the corresponding $r_o$ is computed by using
Eq.(33) and our numerical input $r_s=-0.6$ for $\lambda_o$=1 .
The errors in Table VI
are essentially determined by the observed spurious variation of $\rho_B$
in the range $0\leq \theta \leq 2\pi$.
 As discussed above, this is a numerical
artifact and
 should be considered a systematic effect of the Monte Carlo lattice
simulation.
It is reproduced with three different random number generators, all consistent
with the Kolmogorov-Smirnov test at the level O($10^{-4}$).
At low $J$ this systematic effect
completely dominates the error; the statistical errors, already after 30,000
iterations, are 4-5 times smaller.
\par Table VI also reports the $v_B$ and $Z_{\phi}$ values obtained from the
two-parameter fits to the data using Eq.(29). The resulting $Z_{\phi}$ values
agree well with the one-loop prediction of Eq.(25). As in the one-component
theory,
to perform a more stringent test of the one-loop potential, we have next
constrained $Z_{\phi}$ to its one-loop value in Eq.(25)
and, under such condition, we have done a precise determination
of $v_B$ from a {\it one}-parameter
fit to Eq.(29). This allows a
meaningful comparison with  Eq.(24). In this case the value of
$y_L$ derived from the one-parameter fit to the O(2) lattice data at
$\lambda_o=1.5$ and $r_o=-0.9$ turns out to be $y_L=2.44\pm0.03$ with
${{\chi^2}\over{d.o.f.}}={{3.3}\over{17}}$.
 In Table VII we show
the results of the one-parameter fits to the data at $\lambda_o=$1.0 and 2.0
and the comparison with
Eq.(24) for $y_L=2.44\pm0.03$.
 It is apparent from Table VII
that the one-loop potential reproduces quite well the scaling law of $v_B$
with the bare coupling,
 as previously discovered in the one-component case.
\par The value of $y_L$ obtained from the
lattice simulation of the O(2)-symmetric massless $\lambda\Phi^4$ theory
is $\sim$17$\%$
larger than the value $y_L=2.07\pm0.01$ obtained in the one-component case.
 This is partly due to our choice of fixing the value of
$r_s=-0.6$ at $\lambda_o=1$ with the simple combinatorial
factor $4/3$ discussed above and ignoring finite size
corrections to Eq.(33), see \cite{bra}.
\par With a two-component field, the errors (as estimated from the spurious
$\theta$-dependence of $\rho_B$)
 are larger than in the
one-component theory.
The massless
regime, identified on the basis of the $(J,\rho_B)$
correlation, is found in a range of $r_o$-values
around $r_o=-0.6$ for $\lambda_o=1$. Choosing instead
 $r_s\sim-0.585$ would give $y_L\sim 2.07$.
\par We have analyzed the O(2) lattice data in the same fashion as
already described in the one-component case.
Again, if the agreement with the one-loop formulas
would be a trivial test of perturbation theory
the data should agree at least as well, if not better, with the
leading-log formula based on the perturbative $\beta$-function
\[
 J^{{\rm LL}}(\rho_B)={ {\lambda_o\rho^3_B}\over{1+{{5\lambda_o}
\over{4\pi^2}}
\ln{ { \pi x_{{\rm LL}} }\over{\rho_B}} }}
\]
($x_{{\rm LL}}$ denoting an adjustable parameter to be optimized at each
$\lambda_o$). However, when we fit
the $\lambda_o=$ 1.0, 1.5 and 2.0 data to this formula
we find, respectively,
for 17 degrees
of freedom,
 $(\chi^2)_{{\rm LL}}=$9, 44 and 133 (to
compare with the values $\chi^2=$0.8, 3.3 and 7.3 obtained from the 1-loop
one-parameter fits
 with Eq.(29) when $Z_{\phi}$ is constrained to its value in Eq.(25)).
Note that we are working in a region where the ``$\lambda_o$log'' term is not
small, being of order unity. Thus, the good agreement between the data and
the one-loop formula cannot be due to the higher-order corrections (expected
by perturbation theory) being negligibly small; it must be due to their
complete
{\it absence}. Without
the ``triviality'' argument, this would be an incomprehensible miracle.
\par Finally, to have an idea of the finite size effects, we have repeated the
lattice simulation on a $16^4$ lattice for $\lambda_o=1$ and $r_o=-0.6$.
The computation has been performed at the same
values $\theta=$30, 60, 90, 120, 150, 180, 210, 240
and 270 degrees for the angular source and the data are reported in Table VIII.
 The spurious dependence of the radial field
on $\theta$ is slightly smaller than
in the $10^4$ case and the agreement with the one loop prediction in Eq.(25)
improves.
\par In conclusion, the good agreement between lattice simulation and eqs.
(24,25,29) provides definite evidence that the dependence
of the effective potential on the radial field
in the O(2)-symmetric case is remarkably consistent with the expectations
\cite{resolution,zeit} based on the analysis of the one-component
theory and with the explicit analytical
postgaussian calculation in \cite{rit2} which
numerically confirmed, to very high accuracy, the validity of Eq.(19) in the
O(2) and O(4) case.
\vskip 25 pt
\par  {\bf 5.Conclusions}
\par We have proposed a critical reappraisal of the available results
on the broken phase of $\lambda(\Phi^4)_4$ theory, suggesting a
consistent interpretation of the coexistence of ``triviality'' and
SSB, whose validity is also confirmed by the lattice calculations.
All the existing analytical and numerical evidences can be summarized in the
two
basic points a) and b) illustrated in sect.2. They represent the essential
ingredients for a consistent description of SSB
in pure $\lambda(\Phi^4)_4$ theory. They are naturally discovered in those
approximations to the effective potential which are consistent with
``triviality'' (one-loop, gaussian effective potential, postgaussian
 approximations and so on) where the shifted field(s) appear as generalized
free field(s).
\par As discussed in sect.2,
the perturbative interpretation of the lattice data based on the leading-log
approximation is by no means unique and suffers of internal theoretical
drawbacks. Even more, in sects.3 and 4 we have provided extensive
numerical evidence that in the scaling region, clearly identified with our
procedure, there is no trace of any Higgs-Higgs or Higgs-Goldstone interactions
in the lattice data for $J=J(\phi_B)$.
The {\it measured} effective potential, being remarkably consistent with the
peculiar exponential decay in Eq.(24) and with Eq.(25), is just the sum of the
classical potential and of the zero point energy of a free field.
The corrections, expected on the basis of the leading-log approximation,
are definitely ruled out by our results.
\par The simple scaling laws associated
with the one-loop potential
allow to recover a ``not-entirely trivial'' continuum limit where, despite of
the absence of { \it observable} interactions, one still gets SSB as the
basic ingredient to generate the vector boson masses in the standard model
of electroweak interactions. The $\lambda(\Phi^4)_4$ theory is not ``entirely
trivial''
and can be physically distinguished from a free field theory. Indeed it
exhibits SSB and a first order
 phase transition \cite{resolution,hajj} at a critical temperature $T_c$
depending on the vacuum energy in eq.(16). Also, the origin of SSB from
a classically scale-invariant theory, allows to reconcile the results of
ref.\cite{ped} with ``triviality''.
\par Let us  emphasize, now, that the mass of the Higgs particle does
not represent, by itself, a measure of any interaction and the naive
proportionality between $m_h$ and $\lambda_R$ is not valid outside perturbation
theory. While the latter vanishes the former remains an arbitrary parameter.
As discussed in \cite{con,resolution,zeit,new} substantial
phenomenological implications follow. In our picture, which is consistent with
all
rigorous and numerical results available today, the Higgs particle can be very
heavy and, still, be weakly interacting. In fact, in the absence of the gauge
and Yukawa couplings, it would be trivially free. In the most appealing
theoretical framework, where SSB is generated from a classically
scale-invariant
theory, we get $m^2_h=8\pi^2v^2_R$ so that,
by relating $v_R$ to the Fermi constant (modulo purely electroweak corrections
which are small if the top quark mass is below 200 GeV
\cite{resolution,con,new}
) we get $m_h\sim$ 2 TeV.
In general the Higgs can be lighter or heavier, depending on the size of the
scale-breaking parameter in the classical potential \cite{zeit}. In any case,
beyond perturbation theory, its mass is not related to its decay rate through
$\lambda_R$.
\par  For instance, consider the Higgs decay width to $W$ and $Z$
bosons.  The conventional calculation would give a huge width, of
order $G_F m_h^3 \sim m_h$ for $m_h\sim$ 2 TeV.
 However, in a perturbatively renormalizable-gauge
calculation of the imaginary part of the Higgs self-energy,
this result comes from a diagram in which the Higgs supposedly
couples strongly to a loop of Goldstone bosons with a {\it physical} strength
proportional to its mass squared.
On the other hand, if ``triviality'' is true,
 {\it all} interaction effects of the
pure $\lambda\Phi^4$ sector of the standard model have to be reabsorbed into
two numbers, namely $m_h$ and $v_R$, and there
 are no residual interactions. Thus, beyond perturbation theory,
 that diagram is {\it absent}, leaving a width of order
$g^2 m_h$. Therefore, if ``triviality'' is true, the Higgs is a relatively
narrow
resonance, decaying predominantly to $t \bar{t}$ quarks.
 Although the scalar sector must be
treated non-perturbatively, one may continue to treat the gauge
interactions using perturbation theory.  Effectively, then,
inclusive electroweak processes can be computed as usual, provided
one uses a renormalizable gauge and sets the Higgs self-coupling
and its coupling to the Higgs-Kibble ghosts (the would-be
Goldstones) to zero.  One should avoid the so-called ``unitary
gauge'' where, as well known since a long time \cite{vainst},
there is
no smooth limit $M_w \to 0$ in perturbation theory, as a
consequence of the longitudinal degrees of freedom.
 The possibility
of decoupling the gauge sector from the scalar sector of the standard model
in the limit $g \to 0$ is crucially dependent on the fact that gauge
invariance is carefully retained at any stage. By computing
 in a renormalizable $R_{\xi}$ gauge
one never gets effects (such as cross-sections,
decay rates,...) growing proportionally to the Higgs mass squared {\it unless}
there is a coupling in the theory proportional to the Higgs mass at zero
gauge coupling. Thus, the Higgs phenomenology, on the basis of gauge
invariance,
depends on the details of the pure scalar sector, namely ``triviality'' and
the description of SSB. We believe that our picture, supported as we have shown
by the lattice results for the effective
potential, deserves a very serious attention. The conventional wisdom,
 built up on the perturbative assumption that the Higgs mass is a
measure of its physical interactions, runs a serious risk
 of predicting a whole set
of physical phenomena which exist only in perturbation theory.
\vskip 25 pt
We thank K. Huang and P. M. Stevenson for many useful discussions and
collaboration. We are indebted to P. Cea and L. Cosmai for communication of
their preliminary numerical results.

\vfill
\eject
\begin{tabular}{lccc}
{}~~~$J$ &
{}~~~~~ {$\lambda_o=0.8$~$r_o=-0.36$} &
{}~~~~~ {$\lambda_o=1.0$~$r_o=-0.45$} &
{}~~~~~ {$\lambda_o=1.2$~$r_o=-0.54$} \\
\tableline
-0.700 &
{}~~~~~$-1.0024\pm0.0003$ &
{}~~~~~$-0.9401\pm 0.0003$ &
{}~~~~~$-0.8935\pm0.0003$ \\
\tableline
-0.600 &
{}~~~~~$-0.9540\pm0.0003$ &
{}~~~~~$-0.8950\pm 0.0003$&
{}~~~~~$-0.8512\pm0.0003$ \\
\tableline
-0.500 &
{}~~~~~$-0.8997\pm0.0005$ &
{}~~~~~$-0.8444\pm 0.0003$ &
{}~~~~~$-0.8037\pm0.0003$ \\
\tableline
-0.400 &
{}~~~~~$-0.8371\pm0.0005$ &
{}~~~~~$-0.7867\pm 0.0005$&
{}~~~~~$-0.7493\pm0.0005$ \\
\tableline
-0.300 &
{}~~~~~$-0.7635\pm0.0010$ &
{}~~~~~$-0.7182\pm0.0010$ &
{}~~~~~$-0.6846\pm0.0010$ \\
\tableline
-0.200 &
{}~~~~~$-0.6702\pm0.0010$ &
{}~~~~~$-0.6313\pm0.0010$ &
{}~~~~~$-0.6032\pm0.0010$ \\
\tableline
-0.150 &
{}~~~~~$-0.6112\pm0.0015$ &
{}~~~~~$-0.5764\pm 0.0010$ &
{}~~~~~$-0.5513\pm0.0010$ \\
\tableline
-0.125&
{}~~~~~$-0.5764\pm0.0015$ &
{}~~~~~$-0.5445\pm0.0015$&
{}~~~~~$-0.5213\pm0.0010$ \\
\tableline
-0.100 &
{}~~~~~$-0.5366\pm0.0020$ &
{}~~~~~$-0.5077\pm 0.0020$ &
{}~~~~~$-0.4865\pm0.0024$ \\
\tableline
-0.075 &
{}~~~~~$-0.4902\pm0.0024$ &
{}~~~~~$-0.4641\pm0.0020$ &
{}~~~~~$-0.4459\pm0.0024$ \\
\tableline
-0.050 &
{}~~~~~$-0.4299\pm0.0026$ &
{}~~~~~$-0.4092\pm0.0030$ &
{}~~~~~$-0.3934\pm0.0024$ \\
\tableline
0.050 &
{}~~~~~$0.4273\pm0.0026$ &
{}~~~~~$0.4062\pm0.0030$ &
{}~~~~~$0.3915\pm0.0024$ \\
\tableline
0.075 &
{}~~~~~$0.4877\pm0.0024$ &
{}~~~~~$0.4621\pm 0.0020$ &
{}~~~~~$0.4435\pm0.0024$ \\
\tableline
0.100&
{}~~~~~$0.5348\pm0.0020$ &
{}~~~~~$0.5055\pm0.0020$ &
{}~~~~~$0.4842\pm0.0024$ \\
\tableline
0.125 &
{}~~~~~$0.5749\pm0.0015$ &
{}~~~~~ $0.5428\pm 0.0015$ &
{}~~~~~$0.5201\pm0.0010$ \\
\tableline
0.150&
{}~~~~~$0.6093\pm0.0015$ &
{}~~~~~$0.5754\pm 0.0010$ &
{}~~~~~$0.5505\pm0.0010$ \\
\tableline
0.200 &
{}~~~~~$0.6693\pm0.0010$ &
{}~~~~~ $0.6306\pm 0.0010$ &
{}~~~~~$0.6023\pm0.0010$ \\
\tableline
0.300 &
{}~~~~~$0.7625\pm0.0010$ &
{}~~~~~$0.7172\pm 0.0010$ &
{}~~~~~$0.6838\pm0.0010$ \\
\tableline
0.400 &
{}~~~~~$0.8367\pm0.0005$ &
{}~~~~~$0.7863\pm 0.0005$ &
{}~~~~~$0.7488\pm0.0005$ \\
\tableline
0.500 &
{}~~~~~$0.8993\pm0.0005$ &
{}~~~~~$0.8445\pm 0.0003$ &
{}~~~~~$0.8035\pm0.0003$ \\
\tableline
0.600 &
{}~~~~~$0.9540\pm0.0003$ &
{}~~~~~$0.8951\pm 0.0003$ &
{}~~~~~$0.8513\pm0.0003$ \\
\tableline
0.700 &
{}~~~~~$1.0024\pm0.0003$ &
{}~~~~~$0.9402\pm 0.0003$ &
{}~~~~~$0.8937\pm0.0003$\\
\tableline
{}~~~~~~&
{}~~~~~$Z_{\phi}=33.3\pm 0.6$ &
{}~~~~~$Z_{\phi}=26.0\pm0.4$&
{}~~~~~$Z_{\phi}=21.3\pm0.3$\\
\tableline
{}~~~~~~&
{}~~~~~$v_B=(5.9\pm 2.0)10^{-5}$ &
{}~~~~~$v_B=(6.9\pm 1.4)10^{-4}$&
{}~~~~~$v_B=(3.1\pm0.4)10^{-3}$\\
\tableline
{}~~~~~~&
{}~~~~~$Z^{\rm 1-loop}_{\phi}=32.9$ &
{}~~~~~$Z^{\rm 1-loop}_{\phi}=26.3$&
{}~~~~~$Z^{\rm 1-loop}_{\phi}=21.9$\\
\end{tabular}
\begin{table}
\caption
{ We report the values of $\phi_B(J)$ for the massless case as discussed in
text. At the various  ($\lambda_o$, $r_o$)
we also show the results of the 2-parameter
fit with Eq.(21) and the one loop prediction (25).}
\label{Table I}
\end{table}

\vfill
\eject

\begin{table}
\caption{By using Eq.(21), we show the results of the 1-parameter fits
for $v_B$ at $\lambda_o=$0.8 and 1.2 when
$Z_{\phi}$ is constrained to its one-loop value in Eq.(25). We also show the
predictions from Eq.(24), $(v_B)^{\rm Th}$, for $y_L=2.07\pm0.01$ as determined
from the
fit to the data at $\lambda_o=$1.}
\label{Table II}
\end{table}
\begin{tabular}{cc}
\tableline
{}~~~~~~~~~~{$\lambda_o=0.8~~$} &
{}~~~~~~~~~~{$\lambda_o=1.2~~$} \\
\tableline
{}~~~~~~~~~~{$r_o=-0.36$} &
{}~~~~~~~~~~{$r_o=-0.54$} \\
\tableline
{}~~~~~~~~~~$Z_{\phi}=32.90=fixed$ &
{}~~~~~~~~~~$Z_{\phi}=21.93=fixed$\\
\tableline
{}~~~~~~~~~~$v_B=(7.30\pm0.03)10^{-5}$ &
{}~~~~~~~~~~$v_B=(2.28\pm0.01)10^{-3}$\\
\tableline
{}~~~~~~~~~~~${{\chi^2}\over{d.o.f}}={{5}\over{21}}$&
{}~~~~~~~~~~~${{\chi^2}\over{d.o.f}}={{13}\over{21}}$\\
\tableline
{}~~~~~~~~~~~$(v_B)^{\rm Th}=(7.25\pm0.03)10^{-5}$ &
{}~~~~~~~~~~~$(v_B)^{\rm Th}=(2.29\pm0.01)10^{-3}$
\end{tabular}

\vfill
\eject

\begin{table}
\caption
{ We report the values of $\phi_B(J)$
as obtained by L. Cosmai and P. Cea with their $16^4$ lattice at $\lambda_o=1$
and $r_o=-0.45$. Errors are statistical only. We also report the results of the
fit with Eq.(29) and the one loop prediction (25).}
\label{Table III}
\end{table}
\begin{tabular}{lc}
{}~~~$J$ &~~~~ {$\lambda_o=1.0~r_o=-0.45$}~~~  \\
\tableline
0.100&~~~~~~$0.506086\pm0.91E-04$  \\
\tableline
0.125&~~~~~~$0.543089\pm0.82E-04$ \\
\tableline
0.150&~~~~~~$0.575594\pm0.82E-04$ \\
\tableline
0.200&~~~~~~$0.630715\pm0.62E-04$ \\
\tableline
0.300&~~~~~~$0.717585\pm0.52E-04$ \\
\tableline
0.400&~~~~~~$0.786503\pm0.44E-04$ \\
\tableline
0.500&~~~~~~$0.844473\pm0.41E-04$ \\
\tableline
0.600&~~~~~~$0.894993\pm0.39E-04$ \\
\tableline
0.700&~~~~~~$0.940074\pm0.37E-04$\\
\tableline
{}~~~~~~&${{\chi^2}\over{d.o.f}}={{3.2}\over{9-2}}$\\
\tableline
{}~~~~~~&$Z_{\phi}=26.323\pm0.042$ \\
\tableline
{}~~~~~~&$v_B=(5.783\pm0.013)10^{-4}$\\
\tableline
{}~~~~~~&$Z^{\rm 1-loop}_{\phi}=26.319$ \\
\end{tabular}
\vfill
\eject

\begin{table}
\caption
{We report the one-loop theoretical prediction $J^{\rm{1-loop}}(\phi_B)$
based on Eq.(21),
with the corresponding error $\Delta J$,
which best fits the Monte Carlo data reported in Table III.
$Z_{\phi}=26.319=fixed$ and $v_B=5.7943~10^{-4}$.
 $\chi^2={ { (J^{\rm {1-loop}}-J^{\rm {EXP}})^2}\over{(\Delta J)^2}}$. }
\label{Table IV}
\end{table}
\begin{tabular}{lccc}
\tableline
$\phi_B=0.940074\pm0.37E-04$&~~~$J^{\rm{1-loop}}=0.699975\pm0.82E-04$&
{}~~~$J^{\rm{EXP}}=0.700$&~~~$\chi^2=0.09$\\
\tableline
$\phi_B=0.894993\pm0.39E-04$&~~~$J^{\rm{1-loop}}=0.600001\pm0.78E-04$&
{}~~~$J^{\rm{EXP}}=0.600$&~~~$\chi^2=0.00$\\
\tableline
$\phi_B=0.844473\pm0.41E-04$&~~~$J^{\rm{1-loop}}=0.500042\pm0.72E-04$&
{}~~~$J^{\rm{EXP}}=0.500$&~~~$\chi^2=0.29$\\
\tableline
$\phi_B=0.786503\pm0.44E-04$&~~~$J^{\rm{1-loop}}=0.400027\pm0.67E-04$&
{}~~~$J^{\rm{EXP}}=0.400$&~~~$\chi^2=0.17$\\
\tableline
$\phi_B=0.717585\pm0.52E-04$&~~~$J^{\rm{1-loop}}=0.299952\pm0.65E-04$&
{}~~~$J^{\rm{EXP}}=0.300$&~~~$\chi^2=0.54$\\
\tableline
$\phi_B=0.630715\pm0.62E-04$&~~~$J^{\rm{1-loop}}=0.199982\pm0.59E-04$&
{}~~~$J^{\rm{EXP}}=0.200$&~~~$\chi^2=0.09$\\
\tableline
$\phi_B=0.575594\pm0.82E-04$&~~~$J^{\rm{1-loop}}=0.150010\pm0.64E-04$&
{}~~~$J^{\rm{EXP}}=0.150$&~~~$\chi^2=0.02$\\
\tableline
$\phi_B=0.543089\pm0.82E-04$&~~~$J^{\rm{1-loop}}=0.124943\pm0.57E-04$&
{}~~~$J^{\rm{EXP}}=0.125$&~~~$\chi^2=0.99$\\
\tableline
$\phi_B=0.506086\pm0.91E-04$&~~~$J^{\rm{1-loop}}=0.100062\pm0.54E-04$&
{}~~~$J^{\rm{EXP}}=0.100$&~~~$\chi^2=1.34$\\
\end{tabular}
\vfill
\eject
\begin{table}
\caption
{ The same as in table IV for the leading-log fit.}
\label{Table V}
\end{table}
\begin{tabular}{lccc}
\tableline
$\phi_B=0.940074\pm0.37E-04$&~~~$J^{\rm{LL}}=0.696382\pm0.82E-04$&
{}~~~$J^{\rm{EXP}}=0.700$&~~~$\chi^2=1917$\\
\tableline
$\phi_B=0.894993\pm0.39E-04$&~~~$J^{\rm{LL}}=0.598116\pm0.78E-04$&
{}~~~$J^{\rm{EXP}}=0.600$&~~~$\chi^2=580$\\
\tableline
$\phi_B=0.844473\pm0.41E-04$&~~~$J^{\rm{LL}}=0.499679\pm0.72E-04$&
{}~~~$J^{\rm{EXP}}=0.500$&~~~$\chi^2=19$\\
\tableline
$\phi_B=0.786503\pm0.44E-04$&~~~$J^{\rm{LL}}=0.400981\pm0.67E-04$&
{}~~~$J^{\rm{EXP}}=0.400$&~~~$\chi^2=212$\\
\tableline
$\phi_B=0.717585\pm0.52E-04$&~~~$J^{\rm{LL}}=0.301938\pm0.65E-04$&
{}~~~$J^{\rm{EXP}}=0.300$&~~~$\chi^2=851$\\
\tableline
$\phi_B=0.630715\pm0.62E-04$&~~~$J^{\rm{LL}}=0.202585\pm0.60E-04$&
{}~~~$J^{\rm{EXP}}=0.200$&~~~$\chi^2=1833$\\
\tableline
$\phi_B=0.575594\pm0.82E-04$&~~~$J^{\rm{LL}}=0.152692\pm0.65E-04$&
{}~~~$J^{\rm{EXP}}=0.150$&~~~$\chi^2=1675$\\
\tableline
$\phi_B=0.543089\pm0.82E-04$&~~~$J^{\rm{LL}}=0.127580\pm0.57E-04$&
{}~~~$J^{\rm{EXP}}=0.125$&~~~$\chi^2=1910$\\
\tableline
$\phi_B=0.506086\pm0.91E-04$&~~~$J^{\rm{LL}}=0.102582\pm0.55E-04$&
{}~~~$J^{\rm{EXP}}=0.100$&~~~$\chi^2=2175$\\
\end{tabular}
\vfill
\eject
\begin{table}
\caption
{ The values of $\rho_B(J)$ for the massless case
are reported as discussed in the text. At the various values of $\lambda_o$
and $r_o$ we also show the results of the 2-parameter
fits with Eq.(29) and the one loop prediction (25).}
\label{Table VI}
\end{table}
\begin{tabular}{lccc}
{}~~~$J$ &~~~~ {$\lambda_o=1.0~r_o=-0.6$}~~~ &~~~
 {$\lambda_o=1.5~r_o=-0.9$}
{}~~~ & ~~~
{$\lambda_o=2.0~r_o=-1.2$} \\
\tableline
0.050 &$0.4110\pm0.0130$ & $0.3850\pm 0.0116$
&$0.3753\pm0.0091$ \\
\tableline
0.075 &$0.4686\pm0.0086$ & $0.4337\pm 0.0086$
&$0.4176\pm0.0069$ \\
\tableline
0.100&$0.5133\pm0.0084$ & $0.4724\pm 0.0065$
&$0.4517\pm0.0055$ \\
\tableline
0.125 &$0.5495\pm0.0063$ & $0.5048\pm0.0065$
&$0.4814\pm0.0054$ \\
\tableline
0.150&$0.5819\pm0.0060$ & $0.5332\pm 0.0053$
&$0.5063\pm0.0042$ \\
\tableline
0.200 &$0.6377\pm0.0047$ & $0.5812\pm 0.0040$
&$0.5500\pm0.0037$ \\
\tableline
0.250 &$0.6844\pm0.0039$ &$0.6217\pm0.0034$
&$0.5875\pm0.0029$\\
\tableline
0.300 &$0.7256\pm0.0030$ &$0.6571\pm0.0026$
&$0.6190\pm0.0029$ \\
\tableline
0.350 &$0.7612\pm0.0029$ &$0.6892\pm0.0026$
&$0.6473\pm0.0026$\\
\tableline
0.400 &$0.7938\pm0.0025$ & $0.7178\pm0.0025$
&$0.6731\pm0.0025$ \\
\tableline
0.450 &$0.8246\pm0.0024$ &$0.7435\pm0.0025$
&$0.6969\pm0.0023$\\
\tableline
0.500 &$0.8520\pm0.0023$ &$0.7683\pm0.0021$
&$0.7193\pm0.0023$ \\
\tableline
0.550 &$0.8785\pm0.0021$ &$0.7911\pm0.0021$
&$0.7399\pm0.0018$\\
\tableline
0.600 &$0.9029\pm0.0019$ &$0.8124\pm0.0020$
&$0.7593\pm0.0017$ \\
\tableline
0.650 &$0.9261\pm0.0019$ &$0.8330\pm0.0019$
&$0.7778\pm0.0017$\\
\tableline
0.700 &$0.9481\pm0.0018$ &$0.8520\pm0.0019$
&$0.7953\pm0.0017$\\
\tableline
0.750 &$0.9693\pm0.0018$ &$0.8703\pm0.0017$
&$0.8119\pm0.0015$\\
\tableline
0.800 &$0.9893\pm0.0016$ &$0.8879\pm0.0015$
&$0.8274\pm0.0013$\\
\tableline
{}~~~~~~&$Z_{\phi}=25.0\pm1.6$  &$Z_{\phi}=16.4\pm0.7$
&$Z_{\phi}=12.3\pm0.4$\\
\tableline
{}~~~~~~&$v_B=(1.4^{+1.5}_{-0.9})10^{-3}$  &$v_B=(1.8\pm 0.6)10^{-2}$
&$v_B=(5.6\pm0.8)10^{-2}$\\
\tableline
{}~~~~~~&$Z^{\rm 1-loop}_{\phi}=26.3$ &$Z^{\rm 1-loop}_{\phi}=17.5$
&$Z^{\rm 1-loop}_{\phi}=13.1$\\
\end{tabular}

\vfill
\eject

\begin{table}
\caption{By using Eq.(29), we show the results of the 1-parameter fits
for $v_B$ at $\lambda_o=$1.0 and 2.0 when
$Z_{\phi}$ is constrained to its one-loop value in Eq.(25). We also show the
predictions from Eq.(24), $(v_B)^{\rm Th}$, for $y_L=2.44\pm0.03$ as determined
from the
fit to the data at $\lambda_o=$1.5 .}
\label{Table VII}
\end{table}

\begin{tabular}{cc}
\tableline
{}~~~~~~~~~~~~~~~~{$\lambda_o=1.0~~r_o=-0.6$}~~~~~~~~~~~~~~~~~~~~~ &
{$\lambda_o=2.0~~r_o=-1.2$} \\
\tableline
$Z_{\phi}=26.32=fixed$ &$Z_{\phi}=13.16=fixed$\\
\tableline
$v_B=(7.09\pm0.11)10^{-4}$ &$v_B=(3.79\pm0.03)10^{-2}$\\
\tableline
${{\chi^2}\over{d.o.f}}={{0.8}\over{17}}$&${{\chi^2}\over{d.o.f}}=
{{7.3}\over{17}}$\\
\tableline
$(v_B)^{\rm Th}=(6.85\pm0.09)10^{-4}$ &
$(v_B)^{\rm Th}=(3.89\pm0.05)10^{-2}$
\end{tabular}
\vfill
\eject
\begin{table}
\caption
{ We report the values of $\rho_B(J)$ obtained on our $16^4$ lattice for
$\lambda_o=1.0$ and $r_o=-0.6$. We also show the results of the 2-parameter
fits with Eq.(29) and the one loop prediction (25).}
\label{Table VIII}
\end{table}
\begin{tabular}{lc}
{}~~~$J$ &~~~~ {$\lambda_o=1.0~r_o=-0.6$}~~~  \\
\tableline
0.050&$0.4101\pm0.0098$  \\
\tableline
0.075&$0.4659\pm0.0065$  \\
\tableline
0.100&$0.5105\pm0.0052$  \\
\tableline
0.125 &$0.5470\pm0.0046$ \\
\tableline
0.150&$0.5774\pm0.0035$ \\
\tableline
0.250 &$0.6808\pm0.0023$ \\
\tableline
0.300 &$0.7235\pm0.0022$ \\
\tableline
0.350 &$0.7601\pm0.0022$ \\
\tableline
0.400 &$0.7933\pm0.0022$ \\
\tableline
0.450 &$0.8237\pm0.0020$ \\
\tableline
0.500 &$0.8517\pm0.0019$ \\
\tableline
0.550 &$0.8779\pm0.0018$ \\
\tableline
0.600 &$0.9025\pm0.0018$ \\
\tableline
0.650 &$0.9256\pm0.0014$ \\
\tableline
0.700 &$0.9475\pm0.0013$\\
\tableline
0.750 &$0.9681\pm0.0013$\\
\tableline
0.800 &$0.9874\pm0.0012$\\
\tableline
{}~~~~~~&$Z_{\phi}=26.35\pm1.33$ \\
\tableline
{}~~~~~~&$v_B=(6.8^{+6.3}_{-3.8})10^{-4}$\\
\tableline
{}~~~~~~&$Z^{\rm 1-loop}_{\phi}=26.32$ \\
\end{tabular}
\end{document}